\documentclass[twocolumn,aps,prl,superscriptaddress,amsmath,amssymb]{revtex4}
\usepackage[latin9]{inputenc}
\usepackage{amsmath}
\usepackage{amssymb}
\usepackage{esint}
\usepackage{color}
\usepackage{graphicx}

\makeatletter
\@ifundefined{textcolor}{}
{%
 \definecolor{BLACK}{gray}{0}
 \definecolor{WHITE}{gray}{1}
 \definecolor{RED}{rgb}{1,0,0}
 \definecolor{GREEN}{rgb}{0,1,0}
 \definecolor{BLUE}{rgb}{0,0,1}
 \definecolor{CYAN}{cmyk}{1,0,0,0}
 \definecolor{MAGENTA}{cmyk}{0,1,0,0}
 \definecolor{YELLOW}{cmyk}{0,0,1,0}
}

\usepackage{float}\usepackage{makeidx}\usepackage{CJK}\usepackage{colordvi}\usepackage{color}

\makeatother

\begin{document}
\begin{CJK*}{GB}{gbsn}

\title{Experimental realization of a two-dimensional synthetic spin-orbit coupling in ultracold Fermi gases }

\author{Lianghui Huang}

\affiliation{State Key Laboratory of Quantum Optics and Quantum
Optics Devices, Institute of Opto-Electronics, Shanxi University,
Taiyuan 030006, P.R.China }

\author{Zengming Meng}

\affiliation{State Key Laboratory of Quantum Optics and Quantum
Optics Devices, Institute of Opto-Electronics, Shanxi University,
Taiyuan 030006, P.R.China }

\author{Pengjun Wang}

\affiliation{State Key Laboratory of Quantum Optics and Quantum
Optics Devices, Institute of Opto-Electronics, Shanxi University,
Taiyuan 030006, P.R.China }

\author{Peng Peng}

\affiliation{State Key Laboratory of Quantum Optics and Quantum
Optics Devices, Institute of Opto-Electronics, Shanxi University,
Taiyuan 030006, P.R.China }

\author{ Shao-Liang Zhang}

\affiliation{ Department of Physics, The Chinese University of Hong
Kong, Shatin, New Territories, Hong Kong}

\author{Liangchao Chen}

\affiliation{State Key Laboratory of Quantum Optics and Quantum
Optics Devices, Institute of Opto-Electronics, Shanxi University,
Taiyuan 030006, P.R.China }

\author{Donghao Li}

\affiliation{State Key Laboratory of Quantum Optics and Quantum
Optics Devices, Institute of Opto-Electronics, Shanxi University,
Taiyuan 030006, P.R.China }

\author{Qi Zhou$^{\ddag}$}

\affiliation{ Department of Physics, The Chinese University of Hong
Kong, Shatin, New Territories, Hong Kong}

\author{Jing Zhang$^{\dagger}$}

\affiliation{State Key Laboratory of Quantum Optics and Quantum Optics Devices,
Institute of Opto-Electronics, Shanxi University, Taiyuan 030006,
P.R.China }

\affiliation{Synergetic Innovation Center of Quantum Information and Quantum Physics,
University of Science and Technology of China, Hefei, Anhui 230026,
P. R. China}
\begin{abstract}

\end{abstract}
\maketitle
\end{CJK*}

\textbf{Spin-orbit coupling (SOC) is central to many physical
phenomena, including fine structures of atomic spectra and quantum
topological matters. Whereas SOC is in general fixed in a physical
system, atom-laser interaction provides physicists a unique means to
create and control synthetic SOC for ultracold atoms
\cite{Dalibard}. Though significant experimental progresses have
been made
\cite{spielman,Jing,MIT,Spielman-Fermi,Shuai-PRL,Washington-PRA,Purdue},
a bottleneck in current studies is the lack of a two-dimensional
(2D) synthetic SOC, which is crucial for realizing high-dimensional
topological matters. Here, we report the experimental realization of
2D SOC in ultracold $^{40}$K Fermi gases using three lasers, each of
which dresses one atomic hyperfine spin state. Through spin
injection radio-frequency (rf) spectroscopy \cite{MIT}, we probe the
spin-resolved energy dispersions of dressed atoms, and observe a
highly controllable Dirac point created by the 2D SOC. Our work
paves the way for exploring high-dimensional topological matters in
ultracold atoms using Raman schemes. }

There have been many theoretical proposals in the literature for
creating multi-dimensional synthetic SOC
\cite{Unanyan99PRA,Ruseckas:2005,Juz10PRA,Campbell11PRA,Xu13PRA,AndersonPRL},
so that novel macroscopic quantum phenomena and quantum topological
states could be studied using ultracold atoms
\cite{Stanescu07PRL,Juz08PRL,Vaishnav08PRL,Larson09PRA,Yongping12PRL,Zhang10PRA,Zhu11PRL,Hui11PRL,Yu11PRL}.
Whereas these proposals have not been realized in laboratories,
physicists have also just begun to explore topological phenomena
with ultracold atoms in optical lattices
\cite{Tarruell,Jotzu,Aidelsburger,Miyake}. Here we use the Raman
scheme, to produce a highly controllable 2D synthetic SOC for an
ultracold Fermi gas of $^{40}$K. Such a SOC allows us to create and
manipulate a stable Dirac point on a 2D plane, which is detected by
spin injection rf spectroscopy \cite{MIT}.

\begin{figure}
\centerline{
\includegraphics[width=3in]{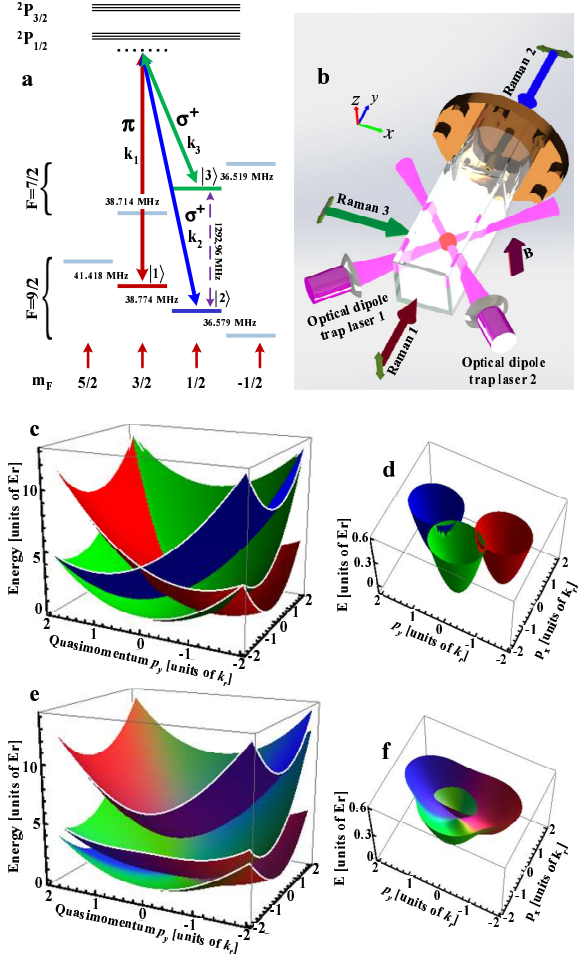}
} \vspace{0.1in}
\caption{\textbf{Two-dimensional synthetic spin-orbit coupling.}
\textbf{a}, Schematic of energy levels of $^{40}$K for creating 2D
SOC. Each of the three Raman lasers dresses one hyperfine spin state
from $|9/2,3/2\rangle$ ($|1\rangle$), $|9/2,1/2\rangle$
($|2\rangle$) and $|7/2,1/2\rangle$ ($|3\rangle$). Atoms are
initially prepared in the free reservoir spin state
$|9/2,5/2\rangle$. \textbf{b}, The experimental geometry and the
laser configuration. The Raman lasers 1 and 2 counter-propagate
along the y axis and the Raman laser 3 propagates along the x axis,
which are linearly polarized along z , x and y directions
respectively. \textbf{c}, Energy-momentum dispersions in the absence
of SOC ( $\Omega_{12}=\Omega_{13}=\Omega_{23}=0$). The three
parabola are displaced from the origin by ${\bf k_{i}}$ respectively
in the $(p_{x},p_{y})$ plane. \textbf{e}, Energy-momentum
dispersions of three dressed states after turning on SOC (
$-\Omega_{12}=2\Omega_{13}=2\Omega_{23}=E_{r}$), which show that the
dispersions of two dressed states touch at a Dirac point. Here,
$\delta_{2}=\delta_{3}=0$. \textbf{d, f} are plots enlarging the low
energy parts of \textbf{c} and \textbf {e}.  In \textbf{e-f }, the
weights of different hyperfine spin states in the dressed ones, the
eigenstates of the Hamiltonian with SOC, are represented by
assigning different colors to the hyperfine spin states. Red, blue
and green represent $|1\rangle$, $|2\rangle$, and $|3\rangle$
respectively. The depth of each color indicates the percentage of
the corresponding spin state in a dressed state. \label{Fig1} }
\end{figure}

We apply three far detuned lasers propagating on the $x-y$ plane to
couple three ground hyperfine spin states, within the $4^{2}S_{1/2}$
ground electronic manifold, $|1\rangle=|F=9/2,m_{F}=3/2\rangle$,
$|2\rangle=|F=9/2,m_{F}=1/2\rangle$ and
$|3\rangle=|F=7/2,m_{F}=1/2\rangle$, where ($F$, $m_{F}$) are the
quantum numbers for hyperfine spin states as shown in Fig. 1a, to
the electronically excited states. Unlike the tripod scheme where a
single excited state is considered \cite{Unanyan99PRA,Ruseckas:2005,Juz10PRA,Stanescu07PRL,
Juz08PRL,Vaishnav08PRL,Larson09PRA}, in $^{40}K$ used here, the
excited states include a fine-structure doublet $4^{2}P_{1/2}$
($D_{1}$ line) and $4^{2}P_{3/2}$ ($D_{2}$ line) with fine structure
splitting of $\sim3.4$ nm. Each of two D-line components
additionally has hyperfine structures. New theoretical
understandings are therefore required.

The microscopic Hamiltonian of our system can be written as
\begin{equation}
\begin{split}
H&=\sum_{i=1}^3(\frac{{\bf
p}^2}{2m}+\varepsilon_{i})|i\rangle\langle
i|+\sum_{j=1}^nE_j|j\rangle\langle j|
\\&+\sum_{i=1}^3\Big(\Omega_ie^{i ({\bf k'}_i\cdot{\bf r}+i\omega_i t+\theta_i)}(\sum_{j=1}^nM_{ji}|j\rangle\langle i|) +h.c.\Big),
\end{split}
\end{equation}
where ${\bf p}$ denotes the momentum of atoms, ${\bf k'_{i}}$
(${|\bf k'_{i}}|=2\pi/\lambda_{i}$) and $\omega_i$ are the wave
vectors and frequencies of three lasers, $\Omega_i$ are the Rabi
frequencies, $i, j$ are the indices for the three ground hyperfine
spin and the excited states respectively, $\varepsilon_{i}$ and
$E_j$ are the ground and excited state energies, n is the total
number of the excited states and $M_{ij}$ is the matrix element of
the dipole transition. Different from the proposal in reference
\cite{Unanyan99PRA,Ruseckas:2005,Stanescu07PRL}, each hyperfine
ground spin state here is dressed by one and only one laser field,
regardless of the excited states it is coupled to. A gauge
transformation, $|i\rangle\rightarrow e^{-i ({\bf k'}_i\cdot{\bf
r}+\theta_i)}|i\rangle$, can be applied to eliminate the phase
$\theta_i$.  All results discussed here are therefore insensitive to
the phase difference, and the sophisticated and challenging phase
lockings are no longer necessary.

Whereas the standard rotating wave approximation gets rid of the
time dependence of the Hamiltonian, for the far detuned lasers, the
excited states can be adiabatically eliminated, and the Hamiltonian
is written as $H_a=p_z^2/(2m)+H_{xy}$,
\begin{equation}
H_{xy}= \sum_{i=1}^3  \left(\frac{({\bf p}-{\bf
k}_{i})^{2}}{2m}+\delta_i\right)|i\rangle\langle i| -\sum_{i'\neq
i}\frac{\Omega_{ii'}}{2}|i\rangle\langle i'|.\label{Hxy}
\end{equation}
Here, $\delta_1$ is set as zero (energy reference) for
simplification, $\delta_2$ ($\delta_3$) corresponds to the
two-photon Raman detuning between Raman laser 1 and 2 (1 and 3), and
${\bf k}_i=\hbar {\bf k}_i'$. All three $\Omega_{ii'}=\Omega_{i'i}$
are real, describing the Raman coupling strength between hyperfine
ground states $|i\rangle\leftrightarrow|i'\rangle$. These values
could be either derived from microscopic calculations (Supplementary
Materials), or measured directly in our experiments \cite{Jing}. The
single-photon recoil momentum $k_{r}=2\pi\hbar/\lambda$ and recoil
energy $E_{r}=k_{r}^{2}/2m$ are taken as natural momentum and energy
units. Since the dispersion along the $z$ direction is not affected
by the lasers, we will focus on the 2D Hamiltonian $H_{xy}$.

Using the two lemmas in Supplementary Materials, we find out that a
doubly degenerate point ${\bf p}_0$ exists in the momentum space,
where ${\bf p}_0$ satisfies two independent equations
\begin{eqnarray}
-\frac{({\bf k}_{1}-{\bf k}_{2})\cdot{\bf p_{0}}}{m}+\delta_{1}-\delta_{2}&=&-\frac{\Omega_{12}\Omega_{13}}{2\Omega_{23}}+\frac{\Omega_{12}\Omega_{23}}{2\Omega_{13}},\nonumber \\
-\frac{({\bf k}_{2}-{\bf k}_{3})\cdot{\bf p_{0}}}{m}+
\delta_{2}-\delta_{3}&=&-
\frac{\Omega_{12}\Omega_{23}}{2\Omega_{13}}+
 \frac{\Omega_{13}\Omega_{23}}{2\Omega_{12}}.\label{ep02}
\end{eqnarray}

Defining the two dressed states with a touching point as a
pseudo-spin-1/2, and projecting the Hamiltonian in equation
(\ref{Hxy}) to this pseudo-spin-1/2 near the degenerate point ${\bf
p}_0$, an effective Hamiltonian at low energies can be obtained straightforwardly,
\begin{equation}
H_{SO}=(\lambda_{x1} p_x+\lambda_{y1} p_y) \sigma_x+(\lambda_{x2} p_x+\lambda_{y2}  p_y)\sigma_z
\end{equation}
where $p_{i=x,y}=p_i-p_{0,i}$. By rotating the momentum and the spin (see Supplementary Material), the Hamiltonian can be simplified as
\begin{equation}
H_{SO}=\lambda_x p_x' \sigma_x'+ \lambda_y p_y'\sigma_z'\label{SOC},
\end{equation}
where both $\lambda_x$ and $\lambda_y$ are finite. Equation
(\ref{SOC}) describes a 2D SOC, which is equivalent to the
Dresselhaus coupling if a simple transformation
$\sigma_x'\rightarrow \sigma_y', \sigma_z'\rightarrow \sigma_x'$ is
applied. Such a 2D SOC directly tells one that the doubly degenerate
point at ${\bf p}_0$ corresponds to a Dirac point with a liner
dispersion at low energies. In particular, both the amplitude and
anisotropy $\lambda_x/\lambda_y$ can be largely tuned (Supplementary
Materials). As a demonstration, figures 1e (1f) and 1c (1d) show the
comparison of the energy-momentum dispersion with and without SOC.
The latter shows that energy dispersions of two dressed states touch
at a Dirac point.

\begin{figure}
\centerline{
\includegraphics[width=3.4in]{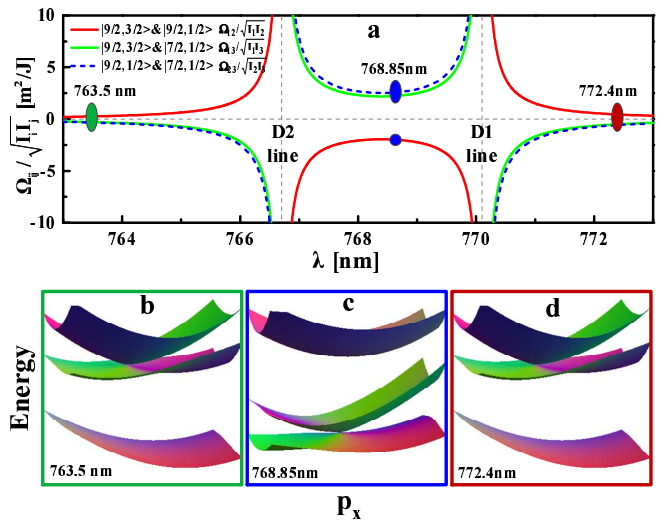}
} \vspace{0.1in}
\caption{ \textbf{Controlling the Dirac point by changing the
wavelength of the Raman lasers.} \textbf{a}, Red, green and blue
curves represent the Raman coupling strength $\Omega_{12}$,
$\Omega_{13}$ and $\Omega_{23}$ as functions of the wavelength
$\lambda$ of the Raman lasers. Depending on $\lambda$,
$\Omega_{12}\Omega_{13}\Omega_{23}$ can be either positive or
negative. $I_i$ is the intensity of the $i$th laser. \textbf{b-d},
The energy bands for the wavelength of 763.5 nm (blue-detuning D2
line), 768.85 nm (between D1 and D2 line) and 772.4 nm (red-detuning
D1 line) respectively. The two higher dispersions touch at the Dirac
point in \textbf{b} and \textbf{d} where
$\Omega_{12}\Omega_{13}\Omega_{23}$ is positive. The two lower
dispersions touch at the Dirac point in \textbf{c} where
$\Omega_{12}\Omega_{13}\Omega_{23}$ is negative. \label{Fig2} }
\end{figure}

In this experiment, a homogeneous bias magnetic field $B_{0}=121.4$
G along the $z$ axis (the gravity direction shown in Fig. 1b)
produces a Zeeman shift to isolate these three hyperfine spin states
from other ones in the Raman transitions, as shown in Fig. 1a. When
three Raman lasers are nearly resonant with three ground states (
$\delta_{i=1,2,3}\approx 0$), the nearest Raman transitions with
other hyperfine states are $|7/2,3/2\rangle\leftrightarrow|2\rangle$
and $|7/2,3/2\rangle\leftrightarrow|3\rangle$, which have large
two-photon Raman detunings about $h\times60$ kHz. Thus we can
neglect other hyperfine spin states and treat this system as a one
with three ground nearly degenerate ground states.

As mentioned before, the Raman coupling strength $\Omega_{ii'}$
includes contributions from all excited states of two D-line
components as shown in Fig. 2a and can be well tuned (Supplementary
Materials). When the wavelength of the Raman lasers is larger than
$D_{1}$ line (770.1 nm) or smaller than $D_{2}$ line (766.7 nm),
$\Omega_{12}\Omega_{13}\Omega_{23}>0$ (see the eigenvalues of
dressed states in Supplementary Materials), and the Dirac point
emerges in the highest two dressed states as shown in Figs. 2b and
2d. In contrast, when the wavelength of the Raman lasers is tuned
between the $D_{1}$ line and $D_{2}$ line,
$\Omega_{12}\Omega_{13}\Omega_{23}<0$ and the Dirac point emerges
when the lowest two dressed states become degenerate, as shown in
Fig. 2c. This is quite different from the work \cite{Yongping12PRL},
where only one excited state is considered and far blue-detuned
lasers are used to make the degenerate dark states the low-lying
ones in the manifold of ground electronic states.

We investigate the energy-momentum dispersions of three dressed
states and the Dirac point by spin injection rf spectroscopy, which
uses rf field to drive the atoms from a free spin-polarized state
into an empty 2D SOC system. We start with a degenerate Fermi gas
${}^{40}$K of $2\times10^{6}$ at the free reservoir spin state
$|9/2,5/2\rangle$ in a crossed optical dipole trap. A homogeneous
bias magnetic field is ramped to $B_{0}=121.4$ G. Then three Raman
lasers are ramped up in 60 ms from zero to its final value.
Subsequently, a Gaussian shape pulse of the rf field is applied for
450 $\mu s$ to drive atoms from the initial reservoir state
$|9/2,5/2\rangle$ to the final empty state with 2D SOC. rf field
does not transfer momentum to the atoms and spin injection occurs
when the frequency of the rf matches the energy difference between
the initial and final states. Since the spin state $|9/2,5/2\rangle$
is coupled via rf to the state $|1\rangle$, rf spectroscopy also
measures the weight of the $|1\rangle$ state, in addition to the
energy dispersions with 2D SOC. Following the spin injection
process, the Raman lasers, the optical trap and the magnetic field
are switched off abruptly, and atoms freely expand for 12 ms in a
magnetic field gradient applied along the x axis. Absorption image
are taken along the z direction. By counting the number of atoms in
state $|1\rangle$ as a function of the momentum and the rf frequency
from the absorption image, we determine the energy band structure
and locate the Dirac point.

\begin{figure}
\centerline{
\includegraphics[width=3in]{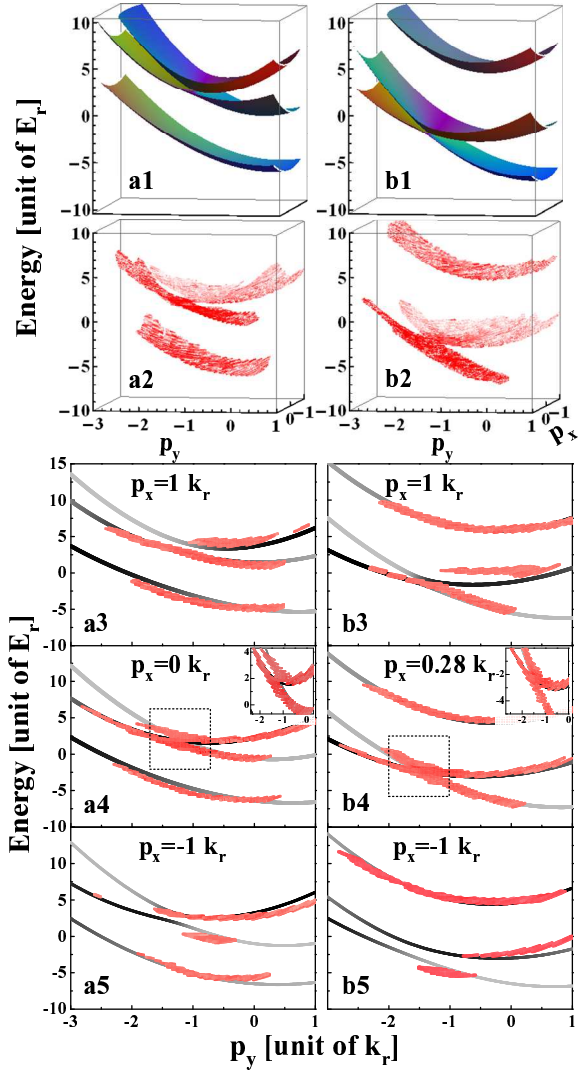}
} \vspace{0.1in}
\caption{ \textbf{ The energy dispersions of dressed atoms measured
by rf spin-injection spectroscopy.} \textbf{a1}-\textbf{a7}, The
one-photon detuning of the Raman lasers is set at the wavelength of
772.4 nm (red-detuning D1). $\Omega_{12}=3.58E_{r}$,
$\Omega_{13}=-3.94E_{r}$, $\Omega_{23}=-4.66E_{r}$,
$\delta_{2}=-5.14E_{r}$, $\delta_{3}=-3.23E_{r}$.
\textbf{b1}-\textbf{b7}, The wavelength is set at 768.85 nm between
the $D_{1}$ line and $D_{2}$ line. $\Omega_{12}=-4.96E_{r}$,
$\Omega_{13}=5.46E_{r}$, $\Omega_{23}=6.46E_{r}$,
$\delta_{2}=-5.2E_{r}$, $\delta_{3}=-2.13E_{r}$. \textbf{a1} and
\textbf{b1} are the theoretical results using the realistic
experimental parameters. \textbf{a2} and \textbf{b2} are
experimental results measured by rf spin-injection spectroscopy.
\textbf{a3}-\textbf{a5}, The cross-section drawings of (a1) and (a2)
in the energy-$p_{y}$ coordinates for different quasimomentum
$p_{x}$ Red dots and solid curves are experimental and theoretical
results respectively. The inset in \textbf{a4} is plot enlarging at
Dirac point. Here, $\lambda_{x1}=0$, $\lambda_{x2}=-0.42k_{r}/m$,
$\lambda_{y1}=1.5k_{r}/m$, and $\lambda_{y2}=-0.07k_{r}/m$ are
obtained by the experimental parameters. \textbf{b3}-\textbf{b5},
The cross-section drawings of (b1) and (b2). The inset in
\textbf{b4} is plot enlarging at Dirac point. $\lambda_{x,1}$,
$\lambda_{x,2}$ $\lambda_{y,1}$ and $\lambda_{y,2}$ are the same as
those in {\bf a}, since $\Omega_{12}/\Omega_{13}$ and
$\Omega_{12}/\Omega_{23}$ remain the same.\label{Fig3} }
\end{figure}

Figure 3(a2) shows the momentum-resolved spin-injection spectrum
when the one-photon detuning of the Raman lasers is set at the
wavelength of 772.4 nm (red-detuning D1) and the corresponding
theoretical plot is shown in Fig. 3(a1). The two higher energy
dispersions touch at a Dirac point as shown in Fig. 3(a2). To
further visualize the Diract point, we plot the energy as a function
of ${\bf p}_y$ for various ${\bf p}_x$, as shown in Fig. 3(a3)-(a5).
When the wavelength of the Raman lasers is tuned to 768.85 nm
between the $D_{1}$ line and $D_{2}$ line, two lower energy
dispersions touch at a Dirac point, as shown in Fig. 3(b1)-(b5).
These two wavelengths were used to investigate the 1D SOC
\cite{Jing,Spielman-Fermi}. For the wavelength of the Raman lasers
at 763.5 nm blue-detuning D2 line, we have observed that two higher
energy dispersions touch at a Dirac point, similar to the 772.4 nm
case. We also perform numerical calculations for the eigenvalues of
the Hamiltonian in equation (\ref{Hxy}) and (\ref{SOC}) according to
experimental parameters and have found out a good agreement between
theory and experiments.

\begin{figure}
\centerline{
\includegraphics[width=2.8in]{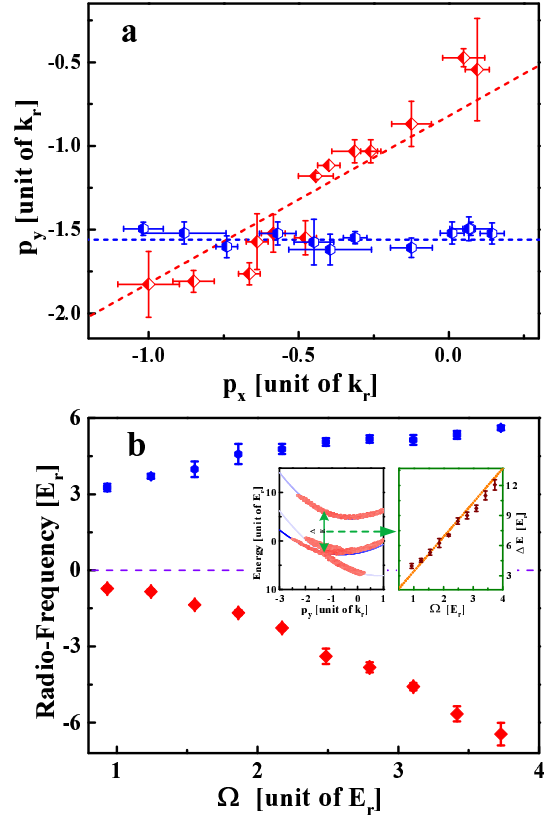}
} \vspace{0.1in}
\caption{ \textbf{Moving the Dirac point.} \textbf{a}, controlling
the Raman detuning $\delta_{2}$ and $\delta_{3}$, the Dirac point
moves along certain trajectories in the momentum space. Blue dots:
$\delta_{2}$ is fixed as $-5.45E_{r}$ and $\delta_{3}$ is tuning at
the range of $[-2E_{r},0.45E_{r}]$. Red dots: $\delta_{3}$ is fixed
as $-0.2E_{r}$ and $\delta_{2}$ is tuning at the range of
$[-7E_{r},-2.5E_{r}]$. The other parameters are
$\Omega_{12}=-4.96E_{r}$, $\Omega_{13}=5.46E_{r}$,
$\Omega_{23}=6.46E_{r}$. \textbf{b}, the energy at the Dirac point
and the corresponding energy of the highest dressed state with the
same momentum ${\bf p}_0$ as a function of the Raman coupling
strength. $-\Omega_{12}=0.91\Omega_{13}=0.77\Omega_{23}=\Omega$,
$\delta_{2}=-5.2E_{r}$ and $\delta_{3}=-2.13E_{r}$. The left inset
in \textbf{b} shows the energy dispersions, The right inset in
\textbf{b} represents the energy separation $\Delta E$ between the
Dirac point and the highest dressed state as a function of the Raman
coupling strength. Here, the one-photon detuning of three Raman
lasers is at the wavelength of 768.85 nm. The error bars represent
the standard deviation of repeated measurements. \label{Fig4} }
\end{figure}

The Dirac point created in this scheme is robust, in the sense that
it moves in the momentum space without opening a gap when
experimental parameters change. This comes from the fact that the
Hamiltonian in equation (\ref{Hxy}) is real, and consequently, the
SOC in equation (\ref{SOC}) cannot contain the $\sigma_y$ term. By
modulating $\delta_{i}$, $\Omega_{i}$ and ${\bf k}_i$, an arbitrary
trajectory of the Dirac point in the momentum space can be designed
in principle. As a demonstration, we fix the Raman detuning
$\delta_{2}$ and measure the positions of the Dirac point on the
$(p_{x},p_{y})$ plane as a function of $\delta_{3}$, as shown in
Fig. 4a (blue dot). When fixing $\delta_{3}$ and changing
$\delta_{2}$, the Dirac point move along a different line. We also
measure the energy at the Dirac point, as well as the energy
separation $\Delta E$ from highest dressed state, as a function of
the Raman coupling strength. Zero energy in Fig. 4b corresponds to
the Zeeman energy splitting between $|9/2,5/2\rangle$ and
$|9/2,3/2\rangle$. The energy of the Dirac point decreases and
$\Delta E$ increases linearly with increasing the Raman coupling
strength (inset of Fig. 4b).

In this work, we have realized a highly controllable 2D SOC and a
stable Dirac point for ultracold fermionic atoms. In the future, we
will develop technologies to open and control the gap at the Dirac
point to generate an effective perpendicular Zeeman field, so that
interesting topological and other exotic superfluids will rise from
s-wave interaction \cite{Zhang10PRA, Zhu11PRL,Hui11PRL,Yu11PRL}. The
heating effect due to spontaneous emission induced by Raman lasers
is neglected here, since we inject atoms to states with SOC that are
initially empty. Such a technical problem will become crucial if
atoms are initially prepared in the states with 2D SOC. It may be
overcome by using atomic species with the absence of the spontaneous
emission, or by exploring the quantum quenching techniques so that
topological properties may be extracted from quantum dynamics in a
relatively short time scale \cite{Sau2015}.

\textbf {Method:}

\textbf {Experimental set-up.} After sympathetic cooling with
bosonic ${}^{87}$Rb in a quadruple-Ioffe configuration magnetic
trap, ${}^{40}$K in the $|9/2,9/2\rangle$ state and ${}^{87}$Rb
atoms in the $|2,2\rangle$ are transferred into an optical dipole
trap formed by two orthogonal 1064 nm laser beams. The Fermi gas is
further evaporatively cooled to $T/T_{F}\approx0.3$ about
$2\times10^{6}$ ${}^{40}$K with ${}^{87}$Rb atoms in the optical
trap ~\cite{four2}, where $T_{F}$ is the Fermi temperature defined
by $T_{F}=(6 N)^{1/3}\hbar\overline{\omega}/k_{B}$, and the
geometric mean of trapping frequencies
$\overline{\omega}\simeq2\pi\times80$ Hz in our system, $N$ is the
number of fermions. After removal of ${}^{87}$Rb atoms, the
fermionic atoms are transferred into the state $|9/2,5/2\rangle$ via
a rapid adiabatic passage induced by a rf field of $80$ ms at $19.6$
G. Here, the transition is addressed by a rf ramp that starts at
6.56 MHz to the end of 6.28 MHz. Then a homogeneous bias magnetic
field along the $z$ axis (gravity direction) is ramped to
$B_{0}=121.4$ G by a pair of coils operating in the Helmholtz
configuration.

\textbf {Raman lasers.} Three Raman lasers are derived from a CW
Ti-sapphire single frequency laser. Two Raman beams 1 and 2 are
frequency-shifted around $+201.144\times2$ MHz and $+220.531\times2$
MHz by two double-pass acousto-optic modulators (AOM), respectively.
The Raman laser 3 is sent through two AOMs with double-pass and
frequency-shifted $-212.975\times4$ MHz. Then three Raman beams are
coupled into three polarization maintaining single-mode fibers
respectively in order to improve stability of the beam pointing and
achieve better beam-profile quality. Behind the fibers, these three
Raman beams have the maximum power $80$ mW for each beam and they
overlap in the atomic cloud with $1/e^{2}$ radii of 200 $\mu m$. The
Raman lasers 1 and 2 counter-propagate along the y axis and the
Raman laser 3 propagates along the x axis, which are linearly
polarized along z , x and y directions respectively, corresponding
to $\pi$, $\sigma$ and $\sigma$ polarization relative to the
quantization axis z as shown in Fig. 1b.

$^{\dagger}$Corresponding author email: jzhang74@sxu.edu.cn,
jzhang74@yahoo.com. $^{\ddag}$ qizhou@phy.cuhk.edu.hk.

\begin{acknowledgments}
We would like to thank Hui Zhai, Shizhong Zhang, Chuanwei Zhang, Han
Pu and Hui Hu for helpful discussions. This research is supported by
the National Basic Research Program of China (Grant No.
2011CB921601), NSFC (Grant No. 11234008, 11361161002, 11222430). QZ
is supported by NSFC/RGC(NCUHK453/13).
\end{acknowledgments}

\begin{widetext}

\section*{SUPPLEMENTARY MATERIAL}

\section*{ 2D synthetic SOC and the Dirac point}

Despite its simple form, equation (2) in the main text leads to
significant results of a highly controllable Dirac point and a 2D
synthetic SOC. To see this fact, we first present two lemmas.
\vspace{0.15in}

{\it Lemma 1}, a real $3\times 3$ matrix
\begin{eqnarray}
\left(
  \begin{array}{ccc}
   -\frac{\Omega_{12}\Omega_{13}}{2\Omega_{23}} & -\frac{\Omega_{12}}{2} & -\frac{\Omega_{13}}{2} \\
   - \frac{\Omega_{12}}{2}  & - \frac{\Omega_{12}\Omega_{23}}{2\Omega_{13}} & -\frac{\Omega_{23}}{2} \\
    -\frac{\Omega_{13}}{2}  & -\frac{\Omega_{23}}{2}  &- \frac{\Omega_{13}\Omega_{23}}{2\Omega_{12}}\\
  \end{array}\label{3by3}
\right),
\end{eqnarray}
has two degenerate eigenstates $|A\rangle$and $|B\rangle$ with zero
eigenenergy  $E_A=E_B=0$,
\begin{eqnarray}
|A\rangle&=&\frac{1}{N_1}\Big(-\Omega_{23}|1\rangle+\Omega_{13}|2\rangle\Big)                 \nonumber\\
|B\rangle&=&\frac{1}{N_2}\Big(\Omega_{13}|1\rangle+\Omega_{23}|2\rangle-\frac{\Omega_{12}(\Omega_{13}^2+\Omega_{23}^2)}{\Omega_{13}\Omega_{23}}|3\rangle\Big)
\end{eqnarray}
where $N_1=\sqrt{\Omega^2_{13}+\Omega^2_{23}}$ and
$N_2=\sqrt{\Omega^2_{13}+\Omega^2_{23}+\frac{\Omega^2_{12}(\Omega^2_{13}+\Omega^2_{23})^2}{\Omega^2_{13}\Omega^2_{23}}}$
are the normalization factors.  The third eigenstate $|C\rangle$ has
the energy
\begin{equation}
E_C=-\frac{1}{2\Omega_{12}\Omega_{13}\Omega_{23}}(\Omega_{12}^{2}\Omega_{13}^{2}+\Omega_{12}^{2}\Omega_{23}^{2}+\Omega_{13}^{2}\Omega_{23}^{2}).
\end{equation}  For positive and negative $\Omega_{12}\Omega_{13}\Omega_{23}$, $|C\rangle$ is the ground or excited state respectively.

\vspace{0.15in}

{\it Lemma 2}, there always exist a finite ${\bf p_{0}}$ to satisfy
a set of three equations
\begin{equation}
\frac{({\bf p_{0}-k_{i}})^{2}}{2m}+\delta_i
=\epsilon-\frac{\Omega_{ii'}\Omega_{ii''}}{2\Omega_{i'i''}},
\,\,\,\,\, i=1,2,3\label{ep0}
\end{equation}
where $\epsilon$ is a constant, $i\neq i'\neq i''$, provided that
$\Omega_{ii'}\neq 0$ for any $i, i'$, and none of these three
momenta ${\bf k}_i=\hbar{\bf k}'_i$  is parallel to other ones  so
that the moentas ${\bf k}_1-{\bf k}_2$, ${\bf k}_2-{\bf k}_3$, and
${\bf k}_1-{\bf k}_3$, two of which are independent, could span a
plane.

These three equations are equivalent to
\begin{eqnarray}
-\frac{{\bf p}_0}{m} \cdot ({\bf k}_1-{\bf k}_2)+(\delta_1-\delta_2)=-\frac{\Omega_{12}\Omega_{13}}{2\Omega_{23}}+\frac{\Omega_{12}\Omega_{23}}{2\Omega_{13}} \label{eq:D1} \\
-\frac{{\bf p}_0}{m} \cdot ({\bf k}_2-{\bf k}_3)+(\delta_2-\delta_3)=-\frac{\Omega_{12}\Omega_{23}}{2\Omega_{13}}+\frac{\Omega_{13}\Omega_{23}}{2\Omega_{12}}\label{eq:D2} \\
-\frac{{\bf p}_0}{m} \cdot ({\bf k}_3-{\bf
k}_1)+(\delta_3-\delta_1)=-\frac{\Omega_{13}\Omega_{23}}{2\Omega_{12}}+\frac{\Omega_{12}\Omega_{13}}{2\Omega_{23}},
\end{eqnarray}
only two of which are independent, since the summation of all three
become zero on both sides of the equation. Thus it is sufficient to
consider only the first two equations.

If the two momenta ${\bf k}_1-{\bf k}_2$ and ${\bf k}_2-{\bf k}_3$
are not parallel to each other, they span a plane. Define this plane
as the $x-y$ plane, the z-component of ${\bf p}_0$ drops off from
these equations, and we could consider a two-dimensional problem.
The solutions of equations \ref{eq:D1} and \ref{eq:D2} form two
lines on this plane. The intersection then uniquely determines the
location of ${\bf p}_0$ on this plane, and the corresponding energy
of this degenerate point is $\epsilon$. Note that regardless of the
microscopic parameters, there is always a trivial solution of ${\bf
p}_0$ at infinity, which is irrelevant to our discussions here.

\vspace{0.08in}

Project the Hamiltonian to the subspace of $|A\rangle$ and
$|B\rangle$, the effective Hamiltonian is,
\begin{equation}
H_e=\left(\begin{array}{cc} \langle A|H|A\rangle & \langle
A|H|B\rangle \\ \langle B|H|A\rangle & \langle B|H|B\rangle
\end{array}\right)=\left(\begin{array}{cc}
\frac{1}{N^2_1}(\Omega^2_{23}\epsilon_{1{\bf
p}}+\Omega^2_{13}\epsilon_{2{\bf p}}) &
\frac{1}{N_1N_2}(\epsilon_{2{\bf p}}-\epsilon_{1{\bf
p}})\Omega_{13}\Omega_{23} \\ \frac{1}{N_1N_2}(\epsilon_{2{\bf
p}}-\epsilon_{1{\bf p}})\Omega_{13}\Omega_{23} &
\frac{1}{N^2_2}\left(\Omega^2_{23}\epsilon_{1{\bf
p}}+\Omega^2_{13}\epsilon_{2{\bf
p}}+\frac{\Omega^2_{12}(\Omega^2_{13}+\Omega^2_{23})^2}{\Omega^2_{13}\Omega^2_{23}}\epsilon_{3{\bf
p}}\right)
\end{array}\right)
\end{equation}
where $\epsilon_{i{\bf p}}={({\bf p-k_{i}})^{2}}/({2m})+\delta_i$. Expand $H_e$ near the degenerate point ${\bf p}_0$, 
and redefine ${\bf p}={\bf p}-{\bf p}_0$, and ${\bf k}_i={\bf
k}_i-{\bf p}_0$, then the effective Hamiltonian can be written as:
\begin{equation}
H_e=\left(\begin{array}{cc} \frac{1}{mN^2_1}(\Omega^2_{23}{\bf
k}_1+\Omega^2_{13}{\bf k}_2)\cdot{\bf p} &
\frac{1}{mN_1N_2}\Omega_{13}\Omega_{23}({\bf k}_1-{\bf
k}_2)\cdot{\bf p} \\ \frac{1}{mN_1N_2}\Omega_{13}\Omega_{23}({\bf
k}_1-{\bf k}_2)\cdot{\bf p} &
\frac{1}{mN^2_2}\left(\Omega^2_{23}{\bf k}_1+\Omega^2_{13}{\bf
k}_{2}+\frac{\Omega^2_{12}(\Omega^2_{13}+\Omega^2_{23})^2}{\Omega^2_{13}\Omega^2_{23}}{\bf
k}_3\right)\cdot{\bf p}
\end{array}\right)+\epsilon\cdot I_{2\times 2}
\label{eq:heff}
\end{equation}
where $I_{2\times 2}$ is the identity matrix.

Dropping the  terms  that are independent on spin,  $H_{e}$ can be
in general written as
\begin{equation}
H_e=(\lambda_{x1} p_x +\lambda_{y1}p_y) \sigma_x+ (\lambda_{x2} p_x
+\lambda_{y2}p_y) \sigma_z
\end{equation}
where the coefficients  $\lambda_{x1}$, $\lambda_{x2}$,
$\lambda_{y1}$, $\lambda_{y2}$ depend on the microscopic parameters
such as ${\bf k}_i$ and $\Omega_{ii'}$.

To simplify the notations, we chose a  coordinate so that
$k_{1y}=k_{2y}$.
Then the effective Hamiltonian (\ref{eq:heff}) can be written as 
\begin{equation}
H_e=ap_x\sigma_x+bp_x\sigma_z+cp_y\sigma_z+\epsilon I
\end{equation}
where $a$, $b$, $c$ and $\epsilon$ are all constant:
\begin{eqnarray}
a&=&\frac{1}{mN_1N_2}\Omega_{13}\Omega_{23}\left(k_{1x}-k_{2x}\right)
                                                       \nonumber\\
b&=&\frac{1}{2mN^2_1}\left(\Omega^2_{23}k_{1x}+\Omega^2_{13}k_{2x}\right)-\frac{1}{2mN^2_2}\left(\Omega^2_{23}k_{1x}+\Omega^2_{13}k_{2x}+\frac{\Omega^2_{12}(\Omega^2_{13}+\Omega^2_{23})^2}{\Omega^2_{13}\Omega^2_{23}}k_{3x}\right)
                                                       \nonumber\\
c&=&\frac{1}{2mN^2_1}\left(\Omega^2_{23}k_{1y}+\Omega^2_{13}k_{2y}\right)-\frac{1}{2mN^2_2}\left(\Omega^2_{23}k_{1y}+\Omega^2_{13}k_{2y}+\frac{\Omega^2_{12}(\Omega^2_{13}+\Omega^2_{23})^2}{\Omega^2_{13}\Omega^2_{23}}k_{3y}\right)=\frac{\Omega^2_{12}(\Omega^2_{13}+\Omega^2_{23})^2}{2mN^2_2\Omega^2_{13}\Omega^2_{23}}(k_{1y}-k_{3y})\nonumber
\end{eqnarray}
When $({\bf k}_1-{\bf k}_2)\times({\bf k}_1-{\bf k}_3)\ne 0$, $a\neq 0$ and $c\neq0$ are satisfied. 

We rotate ${\bf p}$ about the $p_z$ axis by $\varphi$ and the spin
about the $\sigma_y$ axis  by $\theta$. If $\varphi$ and $\theta$
satisfy
\begin{equation}
(a^2+b^2-c^2)\sin2\varphi+2bc\cos2\varphi=0,\,\,\,\,\,\,\,\,(a^2-b^2-c^2)\sin4\theta+2ab\cos4\theta=0,
\end{equation}
the effective Hamiltonian is simplified as
\begin{equation}
H_{SO}=\lambda_x p'_x\sigma'_x+\lambda_y p'_y\sigma'_z,
\end{equation}
 where
\begin{eqnarray}
\lambda_x&=&a\cos\varphi\cos 2\theta-b\cos\varphi\sin
2\theta+c\sin\varphi\sin2\theta
                                      \nonumber\\
\lambda_y &=&a\sin\varphi\sin 2\theta+b\sin\varphi\cos
2\theta+c\cos\varphi\cos2\theta
\end{eqnarray}

A special case is a perfect symmetric configuration,
$\delta_1=\delta_2=\delta_3$, $\Omega_{12}=\Omega_{13}=\Omega_{23}$
so that ${\bf k}_i$ form an equilateral triangle. The Hamiltonian is
$H_{SO}=\frac{d}{2\sqrt{3}m}(p_x\sigma_x+p_y\sigma_z)$ where
$d=|{\bf k}_i-{\bf k}_j|$.

\section*{Calculation of Raman coupling strength}
The Raman coupling strength can be expressed as \cite{two}

\emph{
\begin{gather}\label{1}
\Omega_{ij}=-\frac{I_{0}}{\hbar^{2} c
\epsilon_{0}}\operatorname*{\sum}\limits_{F^{\prime\prime},m^{\prime\prime}_{F}}\frac{\langle{F^{j},
m^{j}_{F}}|er_{q}|{F^{\prime\prime},m^{\prime\prime}_{F}}\rangle\langle{F^{\prime\prime},
m^{\prime\prime}_{F}}|er_{q}|{F^{i},m^{i}_{F}}\rangle}{\Delta},
\end{gather}
} where, $I_{0}=\sqrt{I_{1}\cdot I_{2}}$, and the $I_{i}$ is the
intensity of each Raman laser light. $c$ is the speed of light,
$\epsilon_{0}$ is the permittivity of vacuum, $e$ is the elementary
charge, and $q$ is an index labeling the component of $r$ in the
spherical basis. $\Delta$ is one-photon detuning of Raman lasers.
$|{F^{i},m^{i}_{F}}\rangle$ and $|{F^{j},m^{j}_{F}}\rangle$ are two
ground hyperfine spin states coupled by a pair of Raman laser.
$|{F'',m''_{F}}\rangle$ is the middle excited hyperfine spin state
in the Raman process.

Here, we consider $^{40}K$ atoms, the excited states of the
laser-atom coupling involve a fine-structure doublet $4^{2}P_{1/2}$
($D_{1}$ line) and $4^{2}P_{3/2}$ ($D_{2}$ line) with fine structure
splitting of $\sim3.4$ nm as shown in Fig. \ref{Fig1-supp}. The
Raman coupling strength can be written with two contributions from
$D_{1}$ line and $D_{2}$ line \cite{Lab1}, \emph{
\begin{gather}\label{1}
\Omega_{ij}=-\frac{I_{0}}{\hbar^{2} c
\epsilon_{0}}[\frac{1}{\Delta_{D1}} \times
\operatorname*{\sum}\limits_{F^{\prime\prime}_{1},m^{\prime\prime}_{F}}
\langle{F^{j},
m^{j}_{F}}|er_{q}|{F^{\prime\prime}_{1},m^{\prime\prime}_{F}}\rangle
\langle{F^{\prime\prime}_{1},m^{\prime\prime}_{F}}|er_{q}|{F^{i},m^{i}_{F}\rangle}
\nonumber \\ + \frac{1}{\Delta_{D2}} \times
\operatorname*{\sum}\limits_{F''_{2},m''_{F}} \langle{F^{j},
m^{j}_{F}}|er_{q}|{F^{\prime\prime}_{2},m^{\prime\prime}_{F}}\rangle
\langle{F^{\prime\prime}_{2},m^{\prime\prime}_{F}}|er_{q}|{F^{i},m^{i}_{F}}\rangle].\label{ep02}
\end{gather}
} Here, $\Delta_{D1}$ and $\Delta_{D2}$ are one-photon detuning of
Raman lasers relative to the $D_{1}$ and $D_{2}$ line respectively.
It is required in here that $\Delta_{D1}$ and $\Delta_{D2}$ are very
larger than the hyperfine energy splitting of the excited state.

The Raman coupling strength for $D_{1}$ and $D_{2}$ line can be
simplified as \emph{
\begin{gather}\label{1}
\Omega_{ij}^{D1,(F'',m''_{F})}=-\frac{I_{0}}{\hbar^{2} c \epsilon_{0}} \frac{\langle{F^{j}, m^{j}_{F}}|er_{q}|{F'',m''_{F}}\rangle\langle{F'', m''_{F}}|er_{q}|{F^{i},m^{i}_{F}}\rangle}{\Delta_{D1}} \nonumber \\
=-\frac{I_{0}}{\hbar^{2} c \epsilon_{0}}
\frac{C_{i}^{(F'',m''_{F})}\cdot C_{j}^{(F'',m''_{F})}}{\Delta_{D1}}
\langle{J=1/2}|er_{q}|{J''=1/2}\rangle^{2},\label{ep03}
\end{gather}
} \emph{
\begin{gather}\label{1}
\Omega_{ij}^{D2,(F'',m''_{F})}=-\frac{I_{0}}{\hbar^{2} c \epsilon_{0}} \frac{\langle{F^{j}, m^{j}_{F}}|er_{q}|{F'',m''_{F}}\rangle\langle{F'', m''_{F}}|er_{q}|{F^{i},m^{i}_{F}}\rangle}{\Delta_{D2}} \nonumber \\
=-\frac{I_{0}}{\hbar^{2} c \epsilon_{0}}
\frac{C_{i}^{(F'',m''_{F})}\cdot C_{j}^{(F'',m''_{F})}}{\Delta_{D2}}
\langle{J=1/2}|er_{q}|{J''=3/2}\rangle^{2}.\label{ep04}
\end{gather}
} Here, $C_{i}^{(F'',m''_{F})}$ is the hyperfine dipole matrix
element between the ground $|F^{i},m^{i}_{F}\rangle$ and the excited
hyperfine state $|F^{\prime\prime},m^{\prime\prime}_{F}\rangle$
depending on $\pi$ or $\sigma^{\pm}$ transition, and
$\langle{J=1/2}|er_{q}|{J''=1/2}\rangle$ and
$\langle{J=1/2}|er_{q}|{J''=3/2}\rangle$ are the $D1$ and $D2$
transition dipole matrix element respectively, whose values all can
be found in \cite{three}. Therefore, the Raman coupling strength
$\Omega_{12}$, $\Omega_{13}$ and $\Omega_{23}$ as a function of
one-photon detuning can be obtained from summing contributions from
all the excited hyperfine states of two D-line components using Eqs.
\ref{ep02}-\ref{ep04}, as shown in Fig. 2a in the main text.

\begin{figure}
\centerline{
\includegraphics[width=3.6in]{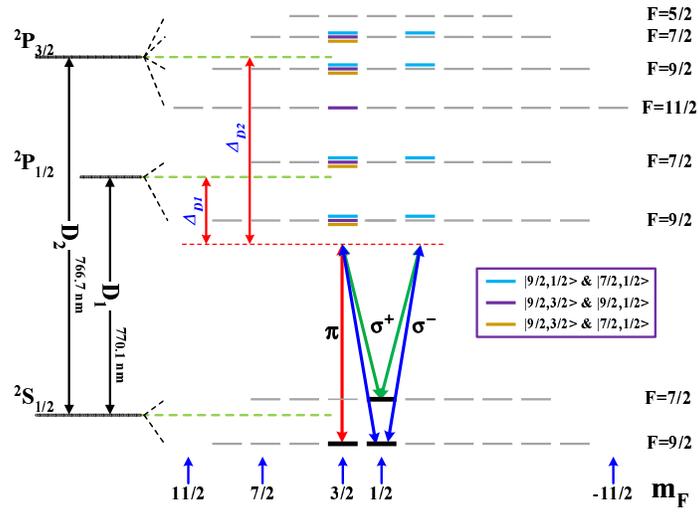}
} \vspace{0.1in}
\caption{(Color online). Schematic of energy levels of $^{40}$K.
\label{Fig1-supp} }
\end{figure}

\end{widetext}

\end{document}